\documentclass{article}

\usepackage{amsmath}
\usepackage{arxiv}
\usepackage[utf8]{inputenc} 
\usepackage[T1]{fontenc}    
\usepackage{hyperref}       
\usepackage{url}            
\usepackage{booktabs}       
\usepackage{amsfonts}       
\usepackage{nicefrac}       
\usepackage{microtype}      
\usepackage{lipsum}		
\usepackage{graphicx}
\usepackage{natbib}
\usepackage{doi}
\setcitestyle{numbers}
\title{Optimizing Time Series Forecasting: A Comparative Study of Adam and Nesterov Accelerated Gradient on LSTM and GRU Networks Using Stock Market Data}

\author{
    Ahmad Makinde \\[0.6 em]  
    Independent Researcher \\[0.6 em]  
    \texttt{ngahmadmak@gmail.com} \\ 
}
\begin{document}
\maketitle

\begin{abstract}
	Several studies have discussed the impact different optimization techniques in the context of time
series forecasting across different architectures. This paper examines the effectiveness of Adam and
Nesterov’s Accelerated Gradient (NAG) optimization techniques on LSTM and GRU neural networks
for time series prediction, specifically stock market time-series. Our study was done by training
LSTM and GRU models with two different optimization techniques - Adam and Nesterov Accelerated
Gradient (NAG), comparing and evaluating their performance on Apple Inc’ closing price data over
the last decade. The GRU model optimized with Adam produced the lowest RMSE, outperforming the
other model-optimizer combinations in both accuracy and convergence speed. The GRU models with
both optimizers outperformed the LSTM models, whilst the Adam optimizer outperformed the NAG
optimizer for both model architectures. The results suggest that GRU models optimized with Adam
are well-suited for practitioners in time-series prediction, more specifically stock price time series
prediction producing accurate and computationally efficient models. The code for the experiments in
this project can be found at \url{https://github.com/AhmadMak/Time-Series-Optimization-Research}
\end{abstract}


\section{Introduction}

The most common algorithm used for training of artificial neural networks is a gradient descent: an optimization algorithm which allows for the minimization a loss function. A local minimum is reached by adjusting model parameters in the direction of the steepest descent. As parameters increase, the minimum value of a function increases in complexity and can even take more than one value. The algorithm begins at the most inefficient parameter sections, taking steps, a local minimum can be reached. But still, standard gradient descent used in the algorithm can converge slowly and be instable, thus advanced optimization methods have been developed.

NAG: Nesterov Accelerated Gradient is one of such methods which was developed by Yurii Nesterov in 1983. By considering what the parameters should be in the next period, NAG enhances the pace of progress; therefore, enhancing responsiveness \cite{nesterov1983method}. Another important method is the Adam optimizer which was created by Kingma and Ba in 2015. Adam uses momentum and adaptive learning rates to allow for faster convergence rates and improved stability. It relies on first and second moment estimates of gradients to update the learning rate for every parameter separately and is extremely relevant when training deep neural networks  \cite{kingma2014adam}.
\\
\\
\\
Convergence speed and stability of learning algorithms are very important in neural networks and particularly time series forecasts. Faster convergence reduces the training time and number of computations needed: which is crucial where constant model updates are required. A stable loss function enables a model to be free from destabilising elements such as gradient explosion or a vanishing gradient: which are detrimental to the learning process    \cite{heaton2018deep}. Aside from this, optimization techniques including the Nesterov Accelerated Gradient and Adam enhance these elements, thus are preferred for many machine learning tasks   \cite{ruder2016overview}.

This study focuses specifically on the application of Nesterov Accelerated Gradient and Adam optimization techniques in time-series forecasting: an area of much focus especially in such areas as finance, weather forecast, health and energy utilization prediction among other applications. The experiments carried out in this study however are only limited to stock market data due to limited computational resources. Dealing with data in the form of stock market time series is complex since time series data tends to be sequential and non-stationary in most cases thus advanced optimization methods must be used to ensure the models generate good forecasts. Earlier works mentioned the successful application of such techniques when enhancing the efficiency of neural networks in predicting time-series, stress their importance in this field  \cite{hewamalage2021recurrent, brownlee2018deep}.The aim of this work is to compare the impact of Nesterov Accelerated Gradient and Adams on stock market data prediction in both GRU and LSTM networks.

\section{Related Works}
\label{sec:headings1}

The ‘Related Works’ section aims to provide a comprehensive overview of works related to our question and is broadly split into 3 overarching sections. The first part aims to summarise key work in relation to the two algorithms assessed in our study: Nesterov Accelerated Gradient and Adams. This is to set the stage for understanding both the significance and impact of these optimization methods in improving convergence speed and stability. The second part of this section will give an overview of the use of neural networks in time series, by discussing different architectures that have been used and the range of problems they have been applied to. The final part will address previous studies comparing optimization techniques used in time-series forecasting by discussing differing methodologies, findings and conclusions.  
\subsection{Nesterov Accelerated Gradient}
Nesterov Accelerated Gradient (NAG) is a momentum-based optimization technique that was introduced by Yurii Nesterov in 1983. NAG considers the future position of the momentum term when calculating the gradient which thus leads to improved convergence properties. The momentum update rule is usually defined as:
\begin{equation}
v_{t+1} = \beta v_{t} + \eta \nabla_{\theta} J(\theta_{t}) 
\end{equation}

\begin{equation}
\theta_{t+1} = \theta_{t} - v_{t+1} \tag{2}
\end{equation}

Where \( v_t \) is the velocity vector, \( \beta \) is the momentum coefficient, \( \eta \) is the learning rate, and \( \nabla_{\theta} J(\theta_t) \) is the gradient of the objective function \( J \) with respect to the parameters \( \theta_t \).

Nesterov’s technique changes the calculation for the gradient by calculating the gradient at the estimated future position of the below parameters: 
\\
\\
\begin{equation}
v_{t+1} = \beta v_t + \eta \nabla_{\theta} J(\theta_t - \beta v_t) 
\end{equation}
\\
\begin{equation}
\theta_{t+1} = \theta_t - v_{t+1} 
\end{equation}
\\
This anticipatory step lowers the oscillations and speeds up convergence, especially when the
optimization landscape is steep and curved. Considering these qualities, Nesterov Accelerated
Gradient has been seen to improve training speed and robustness of neural networks \cite{sutskever2013importance}, \cite{bengio2012advances}.
In the context of neural networks, many studies have shown the power of NAG. For example,
Sutskever et al. \cite{sutskever2013importance} demonstrated that NAG is substantially better than traditional momentum-based
techniques when training deep neural networks, particularly on image recognition tasks. Begino et al.
\cite{bengio2012advances} demonstrated that NAG can avoid low minima and saddle points, thus resulting in better
performance at generalization.
\subsection{Adam}
The Adam (Adaptive Moment Estimation) optimizer, which was introduced by Kingma and Ba in 2015, is a popular optimisation algorithm which benefits from the advantages of both momentum and adaptive learning rates. Adam computes adaptive rates for each of the parameters by keeping running averages of both gradients and their second moments. The update rules for Adam are the following:  
\begin{equation}
m_t = \beta_1 m_{t-1} + (1 - \beta_1) \nabla_\theta J(\theta_t) 
\end{equation}

\begin{equation}
v_t = \beta_2 v_{t-1} + (1 - \beta_2) \left( \nabla_\theta J(\theta_t) \right)^2 \
\end{equation}
Where \( m_t \) and \( v_t \) are estimates for the first and second moments, respectively. Bias-corrected estimates are thus calculated as the following:
\begin{equation}
\hat{m}_t = \frac{m_t}{1 - \beta_1^t} 
\end{equation}

\begin{equation}
\hat{v}_t = \frac{v_t}{1 - \beta_2^t} 
\end{equation}
The parameter update rule is then represented as the following: 
\begin{equation}
\theta_{t+1} = \theta_t - \frac{\eta \hat{m}_t}{\sqrt{\hat{v}_t} + \epsilon}
\end{equation}
With $\eta$ as the learning rate, $\beta_1$ and $\beta_2$ as decay rates for moment estimates, and $\epsilon$ as a small constant to prevent a zero-division error.
\\
Since its introduction, Adam has been studied and applied extensively. Kingma and Ba \cite{kingma2014adam} showed that Adam allows for better performance on many deep learning models including CNNs for image classification, recurrent neural networks for natural language processing and more. Reddi et al. \cite{reddi2019convergence} conducted further studies on Adam’s convergence properties, thus offering both behavioural insights and theoretical guarantees. 

Combining both momentum and adaptive learning rates lets Adam adapt to the shape of the optimization landscape, thus making it able to handle noise and sparse gradients. This has thus allowed for it to become popular in the machine learning community, where it is often the default optimization when training neural networks \cite{kingma2014adam}, \cite{reddi2019convergence}.
\subsection{Overview of Neural Networks used for time-series forecasting}

Neural networks have become crucial for time-series forecasting as they offer advanced models which are able to capture complex temporal dependencies. Two of the most prominent architectures in the field are Long Short-Term Memory (LSTM) networks and Fated Recurrent Units (GRU). 

\subsubsection{Common Architectures  }
\textbf{Long Short-Term Memory (LSTM)}: As introduced in 1997 by Hochreiter and Schmid Huber, they use memory cells that can maintain information over long periods to avoid the issue of vanishing gradient seen in traditional recurrent neural networks (RNNs). LSTM cells feature an input gate, forget gate and output gate which control the information flow. The update equations for LSTMs are the following: 

\begin{equation}
i_t = \sigma(W_i \cdot [h_{t-1}, x_t] + b_i) 
\end{equation}

\begin{equation}
f_t = \sigma(W_f \cdot [h_{t-1}, x_t] + b_f) 
\end{equation}

\begin{equation}
o_t = \sigma(W_o \cdot [h_{t-1}, x_t] + b_o) 
\end{equation}

\begin{equation}
c_t = f_t \cdot c_{t-1} + i_t \cdot \tanh(W_c \cdot [h_{t-1}, x_t] + b_c) 
\end{equation}

\begin{equation}
h_t = o_t \cdot \tanh(c_t) 
\end{equation}

Where \(i_t\), \(f_t\), and \(o_t\) represent the input, forget, and output gates, respectively, and \(c_t\) denotes the cell state, and \(h_t\) denotes the hidden state.

LSTMs have been applied widely to time-series forecasting as they are able to capture dependencies over long ranges. In a study by Gers et al. \cite{gers2003learning}, it was shown that LSTMs were effective when modeling sequential data with long-term dependencies, showing their advantage over conventional RNNs.

\textbf{Gated Recurrent Units (GRU):} GRUs, introduced by Cho et al. \cite{chung2014empirical} in 2014, combine input and forget gates into a single update gate, thus simplifying the architecture and reducing computational complexity whilst still keeping relatively similar performance. The equations for GRU updates are:

\begin{equation}
z_t = \sigma(W_z \cdot [h_{t-1}, x_t] + b_z) \tag{15}
\end{equation}

\begin{equation}
r_t = \sigma(W_r \cdot [h_{t-1}, x_t] + b_r) \tag{16}
\end{equation}

\begin{equation}
\tilde{h}_t = \tanh(W \cdot [r_t \cdot h_{t-1}, x_t] + b) \tag{17}
\end{equation}

\begin{equation}
h_t = (1 - z_t) \cdot h_{t-1} + z_t \cdot \tilde{h}_t \tag{18}
\end{equation}

where \( z_t \) is the update gate, \( r_t \) is the reset gate, and \( \tilde{h}_t \) is the candidate activation. 

GRUs have been seen to perform similarly to LSTMs while needing fewer parameters. A comparative study by Chung et al. \cite{chung2014empirical} corroborated that GRUs are usually as effective as LSTMs in many applications whilst using less computational resources.
\subsubsection{Applications  }
The most common time-series forecasting architectures for time series forecasting are Recurrent Neural Networks (RNNs), of which there are many types including: Encoder-decoder models, Attention mechanisms and most used: LSTMs and GRUs. Both LSTMs and GRUs are often used in time-series forecasting because of an ability to deal with sequential data. Their uses span many domains including financial market prediction, meteorological forecasting and energy consumption forecasting. Financial Market prediction: LSTMs have shown success with stock price prediction because of their ability to model long-term dependencies and temporal relationships in financial data. For example, Qin et al. \cite{qiu2020stock} used LSTM networks when predicting stock prices and showed that LSTMs were able to perform better than traditional learning models regarding prediction accuracy. Their study exemplified the effectiveness of LSTMs when handling volatility and complexity inherent in data from stock markets. 

\textbf{Meteorological Forecasting}: GRUs have been effective in weather forecasting, where an ability to process sequential data with efficiency is important. Rasp and Dubeen \cite{rasp2020weatherbench}, found that GRUs were able to accurately show the dynamics of meteorological data, allowing for increased accuracy in forecasting. Their work shows the ability of GRUs to enhance the precision of weather forecasts with deep learning methods. 
\\
\\
\textbf{Energy Consumption Forecasting}: Kong et al. \cite{kong2019residential} were able to use LSTM networks when predicting residential energy consumption as demonstrated in their study. Their study was able to show the ability to LSTMs to show complex temporal patterns in energy usage data, allowing for more accurate forecasts. In addition, GRUs were able to maintain both high forecasting performance and computational efficiency. 
\subsection{Previous Studies Comparing Optimization Techniques in Time-Series Forecasting }
Extensive literature has explored the effectiveness of different optimization methodologies concerning time-series forecasts. For example, Lim et al. \cite{lim2020survey}presented an analysis of various optimization strategies in stock price prediction with LSTM networks. Their findings revealed that relative to classical gradient descent methods, Adam greatly increased the precision and convergence speed of predictions. Rasp and Dueben \cite{rasp2020weatherbench}also found that Adam provided faster convergence and better generalization compared to NAG, and in case of their weather forecast models converged faster. In another study using energy consumption data, Lai et al. \cite{lai2018modeling}, found that adaptive learning rate-based models (such as Adam or RMSprop) generally outperform static approaches in terms of both speed and accuracy. Despite this there are gaps and limitations of the literature. Many of the studies deal with one type of performance in isolation, and they do not consider that different tasks perform differently with different optimization techniques. In addition to this, there are no well-established benchmarks for comparison across tasks. The purpose of this study is to provide a detailed analysis of the impact Nesterov Accelerated Gradient and Adam optimizers for LSTM and GRU neural networks when predicting stock prices. We focus on this specific application to understand how these optimization techniques perform in a typical forecasting setting.  
costs while still maintaining good levels of computation performance \cite{kong2019residential}. GRUs are great especially for the large-scale forecasting tasks due to their efficiency. The selection of LSTM and GRU in this work is driven by the literature on forecasting which reported their excellent performance for various models. For instance, LSTM is more useful in financial market prediction and energy consumption forecast, whereas GRU are mostly employed for meteorological forecasting \cite{zhou2020energy}
\section{Methodology}
\label{sec:headings2}
\subsection{Neural Network Architectures}
For this study, two popular neural network architectures: Long Short-Term Memory (LSTM) networks and Gated Recurrent Units (GRU) will be used as they are able to capture temporal dependencies well.
\\
\\
\textbf{LSTM Networks}: Introduced by Hochreiter and Schmidhuber, they have proved to handle long-term dependencies well and mitigate the problem of vanishing gradients. Using a complex structure with memory cells and gating mechanisms, they are suitable for problems with intricate sequential patterns \cite{qiu2020stock}.
\\
\\
\textbf{GRU Networks} — a variant of LSTM GRUs are called Gated Recurrent Units and work very similarly to the above LSTMs however instead of having separate input, output, forget gates they combine these three into simplified update gate. This version lowers the computational costs while still maintaining good levels of computation performance \cite{kong2019residential}. GRUs are great especially for the large-scale forecasting tasks due to their efficiency. 
\\
\\
The selection of LSTM and GRU in this work is driven by the literature on forecasting which reported their excellent performance for various models. For instance, LSTM is more useful in financial market prediction and energy consumption forecast, whereas GRU are mostly employed for meteorological forecasting \cite{zhou2020energy}.  

\subsection{Datasets for time-series forecasting}
\subsubsection{Overview of Selected Dataset  }
We will be using a dataset from Yahoo Finance containing the daily closing prices of Apple Inc.’s stock between 2014 and the present month (August 2024), giving us an insight into how the stock’s value has changed over the last decade. It provides us with enough data to examine trends and test how differing optimization strategies and optimizers impact modelling for stock market forecasting. With this dataset, we will be able to examine how Adams and Nesterov optimization algorithms impact prediction accuracy in a real world context .

\subsubsection{Data Preprocessing  }
It is critical to be mindful of data prepossessing when getting datasets ready for training machine learning models, such as neural networks. Normalization, Missing value handling, and Train/Validation/Test Split are the key processes that will be used for the dataset in our study. First, you normalize values, then handle missing data, and lastly, split the processed \& cleaned input into Training, Validation, and Test sets. 
\\
\\
\textbf{Normalization}: Normalization is the process of scaling the features such that they will be in the same range, ensuring that one feature does not disproportionately influence the model because it has a different scale. A common example is normalization, where stock prices are normalized so that the raw values are then given a range to stabilize and speed up learning. In this study, we will also use the Min-Max Scaling formula given by: 
\\
\\
\begin{equation}
x_{\text{norm}} = \frac{x - \min(x)}{\max(x) - \min(x)}
\end{equation}
Where \( x \) represents the original data value, and \( \min(x) \) and \( \max(x) \) are the minimum and maximum values in the dataset, respectively.
\\
\\
\textbf{Handling Missing Values}: One of the most important prepossessing steps in data is to handle missing values, otherwise future steps may be biased or cause wrong models being trained. Different reasons concerning either data collection or corruption may make it impossible to have some values for certain records of information. For numerical data, missing values will be imputed through estimation based on the available data. For continuous variables that are missing, one can use mean or natural interpolation while mode and constant methods are good for categorical or binary data respectively \cite{han2011data} . 

\textbf{Splitting data}: Our data will be split into training, validation and tests sets to evaluate model performance and prevent over-fitting. 
\\
\\
Each of these preprocessing steps are crucial: normalization will prevent learning from being bias by feature with larger scales, handling missing values will prevent inaccuracies arising from incomplete data (allowing for more robust model performance) and finally splitting the data will clearly separate training and evaluation phases, thus enabling accurate assessment of the ability of our models to generalize. 
\\
\\
Overall, proper preprocessing is important in our aim of achieving reliable and high-quality results for our time-series forecasting tasks.
\\
\\
\subsection{Implementation details for Nesterov Accelerated Gradient and Adam}

\textbf{Nesterov Accelerated Gradient (NAG) } 

Nesterov Accelerated Gradient (NAG), as explained in earlier sections, is a variant of gradient descent that uses momentum in order to improve convergence speed and reduce oscillations by looking ahead to compute the gradient at approximate future positions of paraments, helping to adapt the learning direction more effectively.
\\
\\
\textbf{Adam Optimizer}  
Adam (Adaptive Moment Estimation), as described before, is the improved version of AdaGrad and RMSProp. It computes adaptive learning rates for each parameter, borrowing from both worlds, while also adding momentum to speed up convergence. 
\\
\\
Adam is famous for its handling of noisy gradients and quick adaptation of learning rates by using merits of adaptive methods and momentum. Hence, Adam becomes appropriate for non-stationary problems associated with varying data distributions characterized by frequently changing learning rates during the entire training procedure.  

\subsection{Experimental Setup and evaluation metrics}
\subsubsection{Setup  }
The two optimization methods, Adam and Nesterov Accelerated Gradient, used on LSTM and GRU architectures will be compared directly in our experimental framework. For this purpose we utilized Keras library because it has full features, many tools as well as documentation \cite{abadi2016tensorflow}. The latest stable release of Keras for Python 3.12 will serve as the software environment on an online platform based on Google Collab. 
\\
\\
\\
The training parameters are as follows:\begin{itemize}
    \item \textbf{Batch Size:} 1, to maximise granularity in the update of weights during training.
    \item \textbf{Learning Rate:} The default learning rate for the Adam optimizer in Keras, i.e., 0.001, as it offers a good balance between convergence speed and stability.
    \item \textbf{Number of Epochs:} 1, to enable quick initial training.
    \item \textbf{Regularization:} The model was kept simple to assess basic performance; potential overfitting issues were not actively mitigated at this stage.
\end{itemize}

\subsubsection{Evaluation Metrics  }

We use these metrics to evaluate a model: 
\\
\\
\textbf{Convergence Speeds}: Number of epochs that have been met to acceptable loss levels (dataset dependent) quantifies how fast our model learns to make accurate predictions. Wall-clock time will be measured to assess computational efficiency \cite{szegedy2016rethinking}. 
\\
\textbf{Stability}: We will evaluate the consistency of our training process by examining how smooth the loss curve is and whether it exhibits large oscillations. A well-behaved monotonic decrease in loss without significant fluctuations would indicate stable training \cite{bergstra2012random}.
\\
\\
\\
\\
\\
\\
\textbf{Comparison}
\\
\\
To facilitate comparison across datasets and architectures, we shall employ statistical analysis and visualization techniques as follows:
\\
    \textbf{Line Plots}: Display losses against epochs to observe convergence patterns and training instability during the process \cite{bottou2010large}. 
    \\
    \textbf{Convergence Curves}: Measure how fast different optimizers NAG and Adam reach their respective minima by tracking the decline rate of the loss over epochs. 
    \\
    \textbf{Tables}: Contain RMSEs associated with pairs of optimizers and networks. 
\\
\\
    \section{Experiments and Results}

This section contains a comprehensive examination of how LSTM and GRU models perform after being trained on Adam and Nesterov accelerated Gradient (NAG). The assessment emphasizes four elements, which are training loss, validation loss, convergence speed, and Root Mean Square Error (RMSE). Data sets for training, validation, and testing had 1862, 402 and 401 samples, respectively.
\\
\subsection{Training and Validation loss}

\begin{figure}[htbp] 
    \centering
    \includegraphics[width=0.5\linewidth]{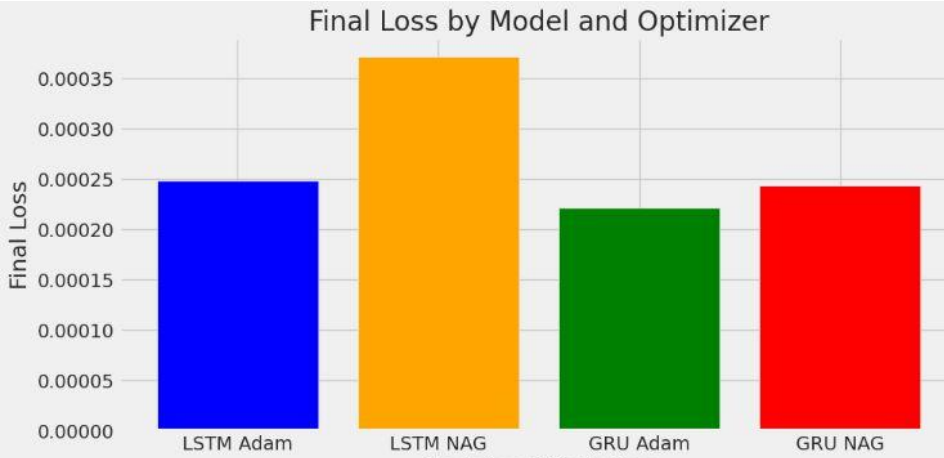}
    \caption{Bar chart showing final loss by model and optimizer.}
    \label{fig:enter-label1}
\end{figure}

10 epochs were used for training each model. Training and validation losses were recorded at every epoch. Final loss values for each model, and optimizer are summarized in \textbf{Figure 1}. For was an LSTM model with Adam, with final training loss of 2.1355 e-04 and a validation loss of 0.0023, the LSTM with NAG had reached a final training loss of 3.4277 e-04 and a validation loss of 0.0031. The GRU models had their peaks at; for Adam, 2.3354e-04 in training and 0.0030 in validation while that of NAG was 2.4583e-04 in training and 0.0012 in validation The results indicate that GRU models performed better in general but particularly with Adam, which consistently had lower validation losses .

\subsection{Convergence Curves}
\begin{figure}[htbp] 
    \centering
    \includegraphics[width=0.5\linewidth]{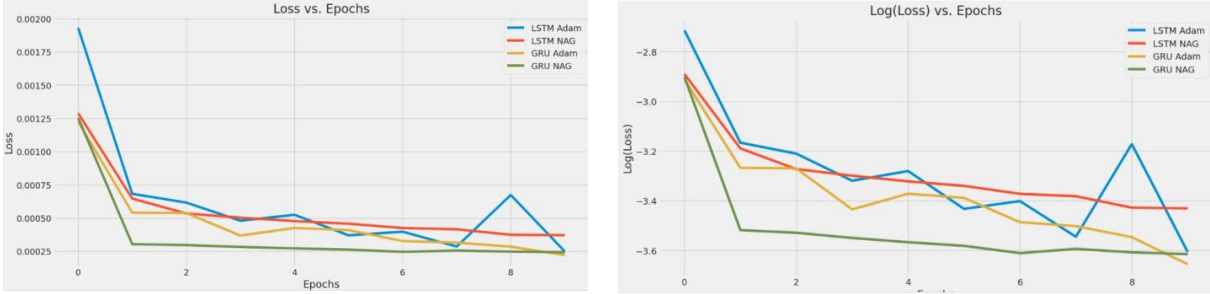}
    \caption{Line chart showing Loss and Log(Loss) vs. Epochs for all 4 models}
    \label{fig:enter-label2}
\end{figure}

Convergence behaviour is depicted in the \textbf{Figure 2}, which show the loss versus epochs and log(loss) versus epochs, respectively. The LSTM models with Adam exhibited a rapid initial reduction in loss but were prone to fluctuations, particularly visible at epoch 8. The NAG optimizer provided more stable convergence, particularly beneficial for the GRU model, which demonstrated a smooth decrease in loss with minimal fluctuations. The log-transformed loss curve emphasizes these stability differences, highlighting the GRU with NAG's superior consistency in reducing loss. The graphs that showcase convergence behaviour are \textbf{Figure 2}, that show loss in relation to epochs and log(loss) against epochs respectively. The LSTM models using Adam had a sharp decrease in their loss function, but this reduction was associated with erratic behaviour especially at epoch 8. On the other hand, the NAG optimizer resulted into more stable convergence which was especially helpful for the GRU model as it illustrated a smooth reduction in loss through slight fluctuations. These stability variations can be emphasized in the log-transformed loss curve where it stands out that GRU with NAG maintains an upper hand over others due to its consistency when it comes to reducing loss.

\subsection{  Comparison of Final Loss by Model and Optimizer}

\begin{figure}
    \centering
    \includegraphics[width=0.5\linewidth]{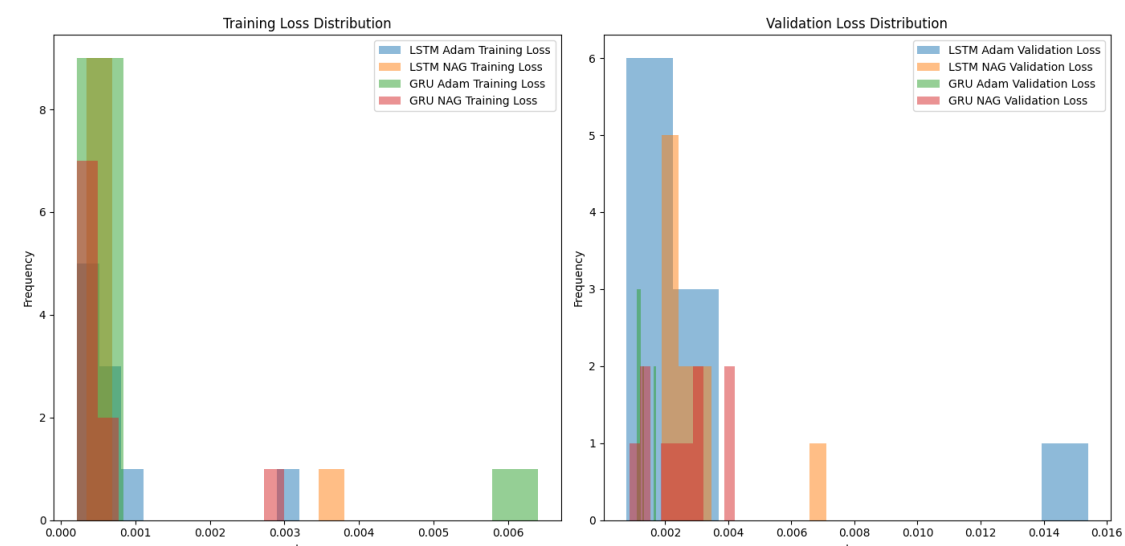}
    \caption{Histograms showing distributions for Final training and validation loss for each model }
    \label{fig:enter-label3}
\end{figure}

\begin{figure}
    \centering
    \includegraphics[width=0.5\linewidth]{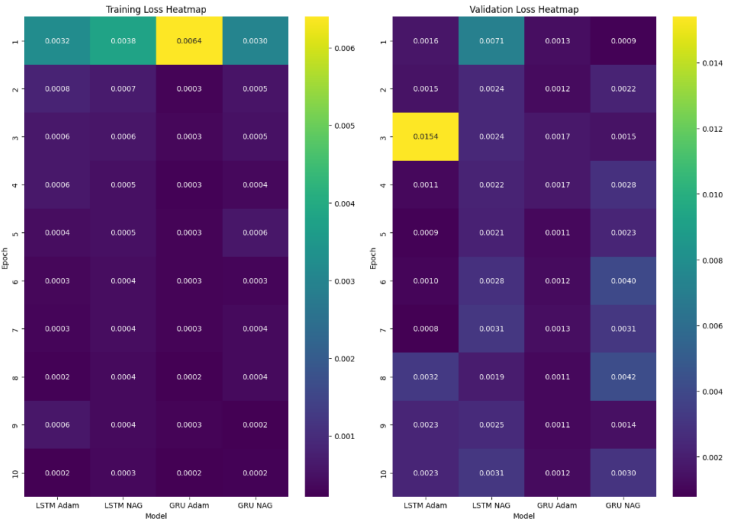}
    \caption{5 Training and Validation Loss Heatmap }
    \label{fig:enter-label4}
\end{figure}

Histograms of training and validation loss distributions are shown in Figure 3 which indicate that GRU models tend to have lower and narrower loss distributions than LSTM ones. The loss values across epochs are also presented in heatmaps for visual comparison between models’ performance in Figure 4. The NAG-GRU had the most stable decrease in loss over time, while other settings displayed varying degrees of stability. These diagrams emphasize the GRU’s better performance during both training and validation phases.
\\

\subsection{  Root Mean Square Error (RMSE) Summary}

\begin{table}[htbp] 
    \centering
    \begin{tabular}{|c|c|} 
        \hline
        Architecture and Optimizer & RMSE \\
        \hline
        LSTM Adam & 176.6259 \\
        LSTM NAG & 187.6416 \\
        GRU Adam & 172.4037 \\
        GRU NAG & 182.2570 \\
        \hline
    \end{tabular}
    \vspace{0.5em} 
    \caption{Table summarising RMSE values for the 4 models}
    \label{tab:my_label}
\end{table}

The RMSE values for each model and optimizer are summarized in the table. Adam’s GRU has the lowest RMSE (172.4037) making it the most accurate predictor. Next is the LSTM with an RMSE of 176.6259 which was achieved by using Adam too. On the other hand, GRU with NAG had an RMSE of 182.2570 whereas the LSTM with NAG recorded the highest RMSE, which was 187.6416. Therefore, these findings show that GRU models especially when used alongside Adam optimizer have increased capacity for predictive accuracy by far when compared to other models.

\subsection{  Summary}

From our experiments, we can conclude that:  
\\

\begin{itemize}
    \item Among the model performance measures, GRU models, particularly those that used Adam optimization, had lower final losses and RMSE than LSTM models.
    \item Regarding their impact as optimizers, it has been observed that the use of Adam as an optimizer always results in decreased values of RMSE and final losses as opposed to NAG, which suggests good overall performance.
    \item The best structure for a model is usually one whose architecture incorporates GRUs working together with Adam optimizers rather than LSTMs.
\end{itemize}

These statements should help in choosing appropriate architectures and optimizers for similar datasets and across prediction problem types like this one. However, other analyses should be done to analyse the impact on larger, more complex datasets.

\section{Discussion}

\subsection{Interpretation of Results}
The findings of this study offer important information about how LSTM and GRU models work when forecasting time-series data from October 2023 by respecting Adam and Nesterov Accelerated Gradient (NAG) optimizers’ training. The best model using GRU with Adam was the one with the lowest RMSE value at 172.4037. It indicates that compared to LSTM, GRU algorithm could catch time-dependencies better because of its simple gating mechanisms in conjunction with an adaptive learning rate function of Adam \cite{kingma2014adam}.

As shown through loss against epochs graph and convergence charts, Adam optimizer enabled faster and more consistent convergence behaviour especially in early stages than NAG optimizer. However, although NAG showed stability during later phases of training, it was outperformed by Adam as far as RMSE is concerned particularly for LSTM model owing to its aggressiveness which makes it predict future gradients thus making updates ineffective within this framework \cite{sutskever2013importance}.

\subsection{Implications for the choice of optimization Technique in Time-Series Forecasting}
The significance of selecting the right optimization method within time-series forecasting is underscored by these findings. Adam being more efficient compared to LSTM and GRU models indicates that it can be used in different types of model structures and is highly effective when dealing with complex, non-linear time series data \cite{cho2014learning}. This seems to suggest that dynamic learning rate adjustment within Adam may help in speeding up convergence while maintaining good generalization exhibited with lower validation losses and RMSEs.

Conversely, for particular reasons including its momentum based approach, the NAG optimizer was less useful in this research. It might not have been beneficial because it tends to overshoot due to an aggressive updating mechanism resulting in terrible learning performance for time-series data as seen with the LSTM model \cite{bengio2012advances}. Therefore such results suggest that practitioners could prefer using Adam particularly when constrained by resource availability where computation power and prediction accuracy are necessary.

\subsection{Limitations of the Study and Potential Areas for Future Research}
The limitations of this study were sparked by lack of sufficient computational resources. The first plan was an empirical one which provided a lot of possibilities for model architecture, hyperparameters scanning and more substantial data sets. However, it had to be restricted because of limited computing power. The capacity constrained the scope of experiments to just a comparative analysis between LSTM and GRU models with Adam and NAG optimizers.

Another limitation was that the data used for training, validation and testing was not very big. The performance evaluation could have been done better if a bigger dataset had been available; this would have also revealed other characteristics about how the models behave on unseen data. The small dataset size might just as well explain why different networks had varied values for their validation losses across optimizers and architectures.

These limitations should be addressed by future research using more computational resources to study a variety of model architectures, including more advanced RNN configurations such as backward and forward LSTM or transformer-based architectures. Besides, the robustness of these findings can be further verified across various data scales by either expanding the dataset or employing techniques of data augmentation \cite{vaswani2017attention}.
In addition, further examination of hybrid optimization methods that combine the best features of Adam and NAG, or new optimizers like RAdam (rectified Adam) or Lookahead may also help to increase convergence speed and accuracy \cite{zhou2020energy}. Another potential research area is how hyperparameter tuning (especially learning rate schedules and batch sizes) impacts model performance which might provide deeper insights into how model architecture interacts with choices made in optimizers and training dynamics thereby influencing effective time-series forecast models’ development.
\section{Conclusion}
The performance of LSTM and GRU models in stock price forecasting was evaluated in this research by looking at how Adam and Nesterov Accelerated Gradient (NAG) affect their results. The finding was that GRU model optimized by Adam had the least RMSE hence it is the best of the combinations experimented with when forecasting stock prices. Thus, it showed that in both LSTM and GRU the most effective optimizer was Adam because it allowed it to converge faster and more steadily.

Due to limited computational resources, the scope of the study was limited to one small dataset, which did not allow thorough exploration of complicated models or larger datasets. However, these findings point out to GRU model with Adam as an emerging prospect for quick and precise stock price prediction. More studies should be conducted that expand on these results in terms of better models and optimizers in future when there are bigger computational resources available. In addition, future research should seek to generalize conclusions across diverse types of time series forecasting including meteorological data, health care data as well as energy consumption data among others.

\pagebreak  
\bibliographystyle{plainnat} 
\bibliography{references} 
\end{document}